# Probing the tunneling site of electrons in strong field enhanced ionization of molecules

J. Wu[1,2], M. Meckel[1], L. Ph. H. Schmidt[1], M. Kunitski[1], S. Voss[1], H. Sann[1], H. Kim[1], T. Jahnke[1], A. Czasch[1], and R. Dörner[1,*]

[1]*Institut für Kernphysik, Goethe-Universität, Max-von-Laue-Strasse 1, D-60438 Frankfurt, Germany.*

[2]*State Key Laboratory of Precision Spectroscopy, East China Normal University, Shanghai 200062, China.*

---

[*] E-mail: doerner@atom.uni-frankfurt.de

**Abstract**

Molecules show a much increased multiple ionization rate in a strong laser field as compared to atoms of similar ionization energy. A widely accepted model attributes this to the action of the joint fields of the adjacent ionic core and the laser on its neighbor inside the same molecule. The underlying physical picture for the enhanced ionization is that the up-field atom that gets ionized. However, this is still debated and remains unproven. Here we report an experimental verification of this long-standing prediction. This is accomplished by probing the two-site double ionization of ArXe, where the instantaneous field direction at the moment of electron release and the emission direction of the correlated ionizing center are measured by detecting the recoil sum- and relative-momenta of the fragment ions. Our results unambiguously prove the intuitive picture of the enhanced multielectron dissociative ionization of molecules and clarify a long-standing controversy.

## Introduction

Understanding and controlling the ionization dynamics of electrons from atoms and molecules in intense ultrashort laser pulses is a primary step towards the coherent control of chemical reactions and photo-biological processes. In the strong-field long-wavelength regime, tunneling is a highly successful concept used to understand the ionization process. In this picture, the effect of the intense light is described as a strong electric field suppressing the atomic binding potential, such that one or multiple electrons can tunnel into the continuum[1]. In the case of multiple ionization of a diatomic molecule, the ionization rate is greatly enhanced if the molecular axis orients parallel to the field as compared to an atom of similar binding energy[2-7].

A widely accepted intuitive mechanism[8-11] explaining this universal behavior is referred to as "enhanced ionization" of the molecular ion after the removal of one or more electrons from the neutral molecule. The two-centre potential binding the outmost remaining electron of a model molecular ion, superimposed with a strong electric field is shown in Fig. 1a. The electron (green blob) is delocalized between the centers. It can tunnel into the continuum (green arrow) through the outer potential barrier. The width and height of the barrier relates to the probability of tunneling. Figure 1b depicts the same model molecular ion with its internuclear distance stretched to a so-called "critical" value. Here, an inner-barrier has emerged between the centers, such that electron population now localizes at each of the cores. Due to this localization, the laser-induced electric field raises the potential energy of the electron attached to the left center ("up-field core"), while lowering it on the right ("down-field core"). To get free, the "up-field" population only needs to tunnel through the inner-barrier between the centers, which is considerably more likely than tunneling through the outer-barrier between the "down-field" center and the continuum. Hence, within this picture, preferential ionization of the "up-field" ion as compared to the "down-field" one is expected, dramatically increasing the overall ionization probability.

Such enhancement of the multiple ionization rate for molecules has been theoretically predicted[8-11] and experimentally observed[2-7] for many years. However, the cornerstone of the underlying physical picture, that it is really the "up-field" atom that gets ionized, stands yet untested. In this respect, some works even contradict[12,13] the predictions of the outlined scenario which is still a puzzle[14,15].

The reason for the missing experimental test of this heart of the enhanced ionization model is twofold. Firstly, in a symmetric single-color laser field, the up-field and down-field directions alternate. It is usually unknown at which swing of the field an electron escapes and the corresponding ion is formed. Equal flux of any charge state to both sides is observed. Secondly, if the symmetry of the electric field is broken either by using a carrier-envelope phase locked few-cycle pulse[16,17] or a two-color pulse[12-15,18], then the asymmetric laser field can potentially redistribute the remaining electrons in the left behind molecular ion. This is a highly interesting phenomenon in itself. Nevertheless, it can overwrite the directionality of the initial enhanced ionization step, prohibiting a conclusive experimental test of the enhanced ionization scenario.

We overcome these hurdles by using a single-color elliptically polarized femtosecond laser pulse. We do not break the symmetry of the laser field. Instead we measure the instantaneous direction of the laser electric field at the moment when the electron escapes for each ionization event. This is achieved by detecting the direction to which the electron is streaked by the elliptically polarized light. We then measure in coincidence if the ion which has released the electron is up- or down-field with respect to this instantaneous field direction.

## Results

**Experimental setup and target.** We choose the van der Waals-force-bound ArXe dimer[19-22] as a clean target to model the stretched diatomic molecule. It has several advantages for probing the enhanced ionization scenario. First, the equilibrium internuclear distance of the neutral dimer ($R_e$~8.0 a.u.)[19,20] is already in the "critical" range, enabling the enhanced ionization process without beforehand stretching. Second, the influence of the molecular orbitals on the ionization is avoided since the electrons are localized at the atomic centers right from the beginning. Third, the ArXe dimer allows us to remove the first electron from the Xe site and then the second electron from the Ar site due to their very different ionization energies[21,22] ($I_p$=11.9 eV for Xe and $I_p$=15.9 eV for Ar). We hence can probe the enhanced ionization by studying the release of the second electron and the ejection direction of $Ar^+$. Although here ArXe dimer is used as a clean system to test the enhanced ionization scenario, the mechanism of enhanced ionization is not special to this system but ubiquitous in molecular ions. The measurements were performed in a reaction microscope of cold target recoil ion momentum spectroscopy (COLTRIMS)[23] (see Methods).

The magnitude and direction of the electric field at the instant of ionization are encoded in the final, detectable momentum vector of the electron. The rotating electric field of an elliptically polarized pulse leads to a final electron momentum approximately perpendicular to the instantaneous field vector due to "angular streaking"[24-30]. Hence, the field direction at the instant of ionization is mapped to the momentum direction of the released electron. For illustration purposes, as sketched in Fig. 1c, if an electron is freed when the laser field vector points along +$y$, its final momentum ($p_e$) will be along -$z$. Likewise, an instantaneous field along –$y$ will lead to a final electron momentum along +$z$. Therefore, any possible asymmetry in ionization rate of the ArXe oriented along the $y$-axis (up/down in Fig. 1c) will be turned into an asymmetry of the electron momentum distribution along the $z$-axis (left/right in Fig. 1c). Technically, instead of detecting electron momentum directly, we detect the momentum vector of the center of mass of the Coulomb exploding $Ar^+Xe^+$ ion. By momentum conservation, this center-of-mass or sum-momentum of the recoil ions ($p_{sum}$) is equal and opposite to the sum-momentum of the two emitted electrons. Meanwhile, in addition to this center-of-mass motion of the molecular ion, the ionic cores repel each other, yielding a much bigger relative momentum ($p_{rel}$) along the former molecular axis. Hence we obtain the information of the direction of the emitted electrons in the molecular frame of reference by only detecting in coincidence the 3D-momenta of both fragment ions.

**Ion momentum distributions of doubly ionized ArXe.** We focus on the Coulomb exploding double ionization channel $Ar^+$+$Xe^+$. The elliptical light suppresses recollision of the freed electrons, thus the two electrons escape sequentially[27] during the laser pulse. As the laser field is elliptical, and the tunneling probability depends steeply on the field strength, each ionization step will maximize when the field vector is along the major polarization axis, yielding two burst of ionization per cycle. As depicted in Fig. 2, we find that double ionization maximizes for molecules lined up with the major axis. We restrict our data analysis and the discussion to such molecules by selecting only fragmentations within a cone of 45° around the major polarization axis.

The first ionization step of removing an electron from Xe will necessarily be symmetric, i.e. ionization when the laser field points in the +$y$ and –$y$ directions has equal probability, yielding symmetric momentum distribution of the first electron along the $z$-axis. No field direction dependent ionization rate of the first electron is expected since it starts from the neutral molecule. The second step of ejecting an electron from the neutral Ar with the neighboring $Xe^+$ now is the one for which we test the enhanced ionization scenario. A preferred directionality of this second electron will lead to an asymmetry in the ion sum-momentum ($pz_{sum}=pz_{Ar+} + pz_{Xe+}$) distribution. If we do not distinguish between emission of $Ar^+$ to the $+y$ or $-y$ directions, the $pz_{sum}$ for $Ar^+ + Xe^+$ is symmetric (gray squares in Fig. 3a). It is the convolution of the momentum distributions of two sequentially removed electrons. As first and second electron can both be emitted either in $+y$ or $-y$ direction, the combination yields four possible sum-momenta[27,28,31] as indicated with the orange bars in Fig. 3a (see Methods). The solid gray curve shows a fit of the convolution by assuming a Gaussian distribution of each electron momentum (see Methods), which returns $pz_{e1}$=0.45±0.07 and $pz_{e2}$=0.61±0.02 a.u. in agreement with the fact that the second electron from Ar is released by a higher laser intensity in the laser pulse than the first one from Xe.

**Discussion**

We will now address the essential question whether or not the second ionization step happens at the up-field core, as predicted by the classical model of enhanced ionization. We break the symmetry of the problem by gating on the orientation of the $Ar^+ + Xe^+$ breakup. In Fig. 3a, the red squares show an enhancement in rate at positive $pz_{sum}$ when the $Ar^+$ departs along $+y$. This shows that the $Ar^+$ ion is born on the up-field site of the molecule in our clockwise light polarization. The relative-momentum ($p_{rel} = p_{Ar+} - p_{Xe+}$) distribution in Fig. 3b very intuitively shows this behavior when we select events where the field at the release of the second electron pointed to $+y$ by gating on $pz_{sum}>1.0$ a.u. for clockwise elliptical light. It shows a strong orientation of the $Ar^+ + Xe^+$ breakup with the $Ar^+$ being up-field. The same curves are recorded with the opposite light helicity, i.e. anticlockwise polarization. The results shown in Figs. 3c,d are mirror images of the clockwise case, confirming the validity of our data. All these results conclusively show that the second electron is more favorable to be freed from the up-field potential well.

To obtain the relative emission probabilities of the second electron from the up-field and down-field potential wells, we fit the asymmetric $pz_{sum}$ distributions with the same model used above, but now featuring two separate fitting factors $A_{2+}$ and $A_{2-}$ for the magnitudes of the $pz_{e2}>0$ and $pz_{e2}<0$ cases instead of a common one (see Methods). The fits (red and blue curves in Fig. 3a) reveal that the probability for the second electron being released from the up-field potential well is about ~2.41±0.16 higher than that from the down-field potential well.

In conclusion, we gave a direct proof that enhanced multielectron ionization of a diatomic molecule is indeed more likely to happen at the up-field than at the down-field centre. This result clarifies the long-time controversy on the electron localization assisted multiple ionization of molecules.

**Methods**

**Experimental technique.** We performed experiments with femtosecond laser pulses (35 fs, 790 nm, 8 kHz) produced from a multipass amplifier Ti:sapphire laser system (KMLabs Dragon). The laser pulses were sent into a standard cold target recoil ion momentum spectroscopy (COLTRIMS)[23] setup and focused by a concave mirror with a focal length of 7.5 cm onto a supersonic gas jet. A mixture of rare gases of Ar and Xe with a ratio of 7:1 and driving pressure of 3.5 bar is used to generate ArXe dimer after its expansion through a 30 μm nozzle. As illustrated in Fig. 1c, the light and molecular beam propagate along the $-t$ and $y$-axes, respectively. The photo-ionization created ions were accelerated with a weak (~14.7

V/cm) static electric field and detected by a time and position sensitive micro channel plate detector[32] at the end of the spectrometer.

The major axis of the elliptically polarized light was fixed to be along the *y*-axis. The recoil momentum from the ejected electrons onto the ions could be clearly distinguished along the minor axis of the polarization ellipse, i.e. *z*-axis or time-of-flight direction of our spectrometer where we have the best momentum resolution. For input laser pulse linearly polarized along the *y*-axis, the handedness and ellipticity of the ellipse was controlled by varying the relative orientation of the fast-axis of a half-wave plate which was put in front of a quarter-wave plate before the vacuum chamber. The fast-axis of the quarter-wave plate was fixed to be along *y*-axis. This defined the major axis of the elliptically polarized light. The peak intensity and ellipticity of the pulse were measured to be $I_0 \sim 4.3 \times 10^{14}$ W/cm$^2$ and $\varepsilon \sim 0.68$, respectively.

**Two-particle correlation identifies ArXe isotopes.** As shown in Fig. 4a, there are seven Xe isotopes which can form ArXe dimers in our gas jet. The sharp stripes displayed in Fig. 4b due to the ion-ion time-of-flight correlation allow us to identify and distinguish the ArXe isotopes. All the isotopes show similar behaviors, we use the Ar$^{134}$Xe and Ar$^{136}$Xe for the discussion in this work since they are best separated from the others.

**Convolution model of the ion sum-momentum distribution.** Since the recollision process is mostly suppressed in the elliptically polarized pulse, the two electrons are predominantly ejected sequentially [27]. As the tunneling probability strongly depends on the field strength, the ionization steps maximize when the laser field is orientated along the major axis (see Fig. 2), yielding streaked electron momentum along the polarization minor axis (see Fig. 1c). Due to momentum conservation, the sum of all electron momentum vectors is the inverse of the sum of all ion momentum vectors. Therefore, by measuring the latter, we can determine the former.

The sum-momentum distribution of the sequentially released two electrons can be modeled as the convolution of them. As illustrated in Fig. 5, each individual electron freed when the laser field vector parallel to the *y*-axis results in two peaks centered at $\pm pz_{e1}$ (Fig. 5a) or $\pm pz_{e2}$ (Fig. 5b) along the *z*-axis. The convolution of them yields four possible momenta (Fig. 5c). The central peaks, $pz_{e1} - pz_{e2}$ and $pz_{e2} - pz_{e1}$, are filled with events that two electrons are freed when the laser field points to opposite directions (back-to-back). The side peaks, $pz_{e1} + pz_{e2}$ and $-pz_{e1} - pz_{e2}$, account for the cases that both electrons are released by the laser field along the same direction. In more details, by assuming a Gaussian shape of the electron momentum distribution, i.e. $A_k/[\sigma_k \sqrt{(2\pi)}] \times \exp[-0.5(pz-pz_{ek})^2/\sigma_k^2]$ (k=1 or 2 accounts for the first or second electron), the recoil ion sum-momentum (i.e. the reverse of sum-momentum of the sequentially freed two electrons) can be expressed as[28]

$$pz_{sum12}(pz) = \frac{1}{\sqrt{2\pi}} \sum_{i,j=+,-} \frac{A_{12ij}}{\sigma_{12ij}} \exp[-0.5(pz - pz_{sumij})^2 / \sigma_{12ij}^2]$$

where $A_{12ij}=A_{1i} \times A_{2j}$, $\sigma^2_{12ij}=\sigma^2_{1i} + \sigma^2_{2j}$, and $pz_{sumij}= -(ipz_{e1} + jpz_{e2})$ accounts for four possible ion sum-momenta.

We fit the measured ion sum-momentum distribution in Fig. 3 by using the above equation of $pz_{sum12}$. For the symmetric case (gray curve in Fig. 3a), the emission probabilities of the electrons by laser field pointing to +*y* and –*y* are assumed to be the same, i.e. $A_{1+}= A_{1-}$ and $A_{2+}= A_{2-}$. For the asymmetric cases with a gate on the emission direction one of the fragment ions (red and blue curves in Figs. 3a,c), the second electron is assumed to have different probabilities to left and right, i.e. $A_{2+} \neq A_{2-}$.

**Acknowledgments** We acknowledge conclusive discussions with C.L. Cocke, R.R. Jones and A. Becker. The work was supported by a Koselleck Project of the Deutsche Forschungsgemeinschaft. J.W. acknowledges support by the Alexander von Humboldt Foundation.

**Author Contributions** J.W. designed the experiments and performed the measurements and data analysis; M.M., L.S., M.K., S.V., H.S., H.K., T.J., A.C., and R.D. provided technique support during the measurements and data analysis; J.W., M.M., and R.D. wrote the paper.

**Additional information**

**Competing financial interests**: The authors declare no competing financial interests.

**Reprints and permission** information is available online at http://npg.nature.com/reprintsandpermissions/

Correspondence and requests for materials should be addressed to R.D. (email: doerner@atom.uni-frankfurt.de).

## Figure Legends

**Figure 1**
**Enhanced ionization scenario and the angular streaking concept. (a, b)** Field-dressed potentials (orange surfaces) of a model diatomic molecule ion at **(a)** short and **(b)** large (critical) internuclear distances. In the latter case, the electron (green blob) nonadiabatically localized in the up-field potential well is more favorable to be freed through the narrow and low inner-barrier when the molecule stretches to the critical internuclear distance. **(c)** Sketch of the angular streaking concept and our coordinate system. Due to the streaking of the elliptically polarized laser pulse (red helix), an electron will receive a final momentum (green arrows) oriented perpendicular to the laser field (violet arrows) at its tunneling. The instantaneous laser field vector at the moment of ionization hence can be retrieved from the streaked electron momentum $p_e$ (green arrows) or its recoil on the ion sum-momentum $p_{sum}$ (orange arrows). The olive helices in the dashed insets show sample momentum trajectories of the driven electron during the pulse. The ejection direction of the ionizing center or molecular orientation (orange blobs) is derived from the relative-momentum between the fragment ions $p_{rel}$ (blue arrows).

**Figure 2**
**Relative momentum and angular distributions of $Ar^+ + Xe^+$ breakup**. **(a)** Measured relative momentum distribution between $Ar^+$ and $Xe^+$ from Coulomb exploding double ionization of ArXe using elliptically polarized femtosecond laser pulses whose major and minor axes orientated along $y$- and $z$-axis, respectively, as sketched by the red ellipse in **(b)**. **(b)** The corresponding angular distribution of the $Ar^+ + Xe^+$ breakup in the polarization plane of the laser pulse. The double ionization rate maximizes for molecules orientated along the polarization major axis of our elliptically polarized laser field.

**Figure 3**
**Ion sum- and relative-momentum distributions of $Ar^+ + Xe^+$ breakup. (a,c)** Ion sum-momentum distributions. These reflect the recoil of two sequentially emitted electrons. The red squares and blue circles are events where $Ar^+$ fly in the $+y$ and $-y$ directions, and the gray squares are the sum of them. To guide the eye, four possible sum-momenta of the sequentially ionized electrons are indicated by the pink bars in **(a)** under the gray curve. **(b,d)** Ion relative-momentum distributions. These reflect the ArXe orientation. Only the events when the instantaneous field vector (violet arrow) at the release of the second electron pointing along **(b)** $+y$, or **(d)** $-y$ are selected by choosing $pz_{sum}>1.0$ a.u. (see text for details). In agreement with the intuitive picture of the enhanced ionization, the second electron from ArXe is favored to be freed when the instantaneous laser field is pointing from $Xe^+$ towards the Ar atom. As sketched by the inset red ellipses at the top-right corners of **(a)** and **(c)**, the sense of rotation of the electric field vector of our laser pulse is adjusted to be **(a,b)** clockwise, and **(c,d)** anticlockwise, respectively.

**Figure 4**
**Two-particle correlation identifies ArXe isotopes. (a)** Measured times-of-flight mass spectrum of the $Xe^+$ isotopes. **(b)** Correlations between the times-of-flight of the Coulomb exploding $Ar^+$ and $Xe^+$ following two-side double ionization of ArXe. Each sharp stripe stands for one ArXe isotope.

**Figure 5**
**Schematic illustration of the momentum convolution of two sequentially released electrons.** The orange bars indicate the momentum peaks of the **(a)** first and **(b)** second electrons, and **(c)** the eventually measured distribution as a result of their convolution.

**Figure 1**

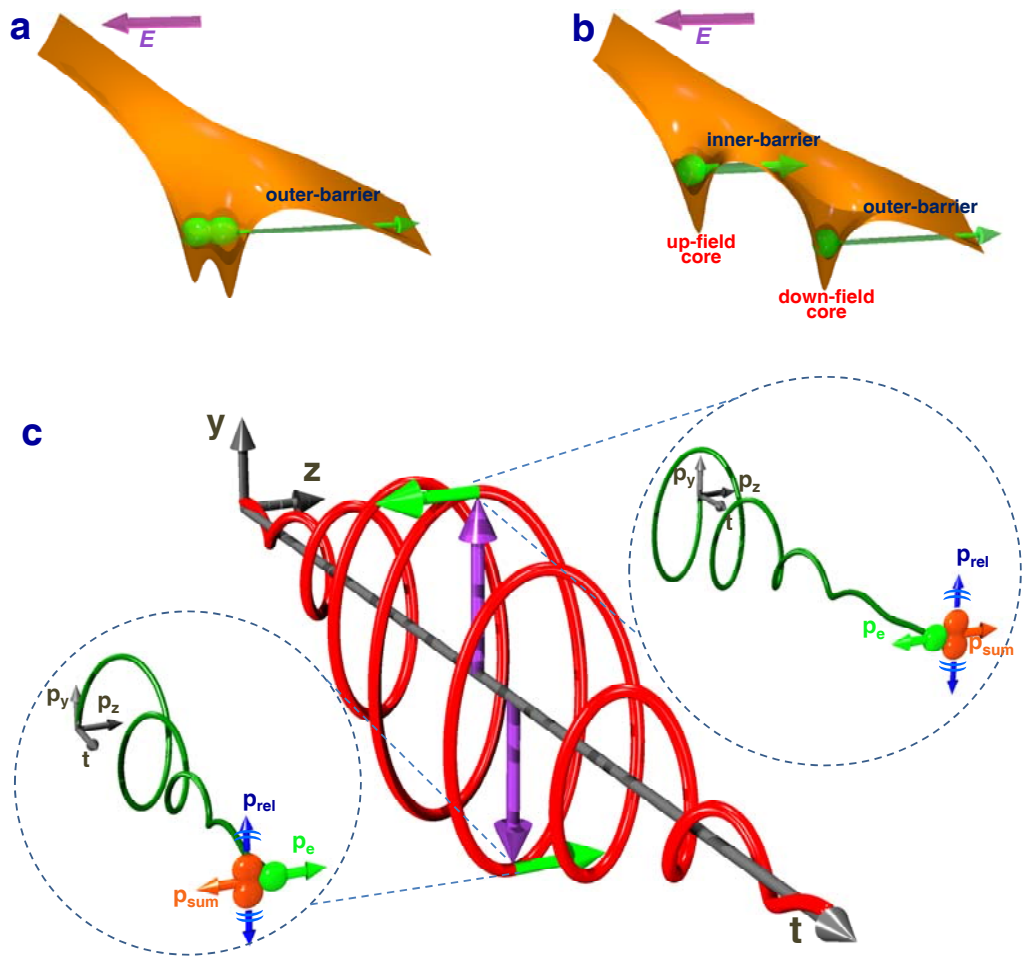

**Figure 1: Enhanced ionization scenario and the angular streaking concept. (a, b)** Field-dressed potentials (orange surfaces) of a model diatomic molecule ion at **(a)** short and **(b)** large (critical) internuclear distances. In the latter case, the electron (green blob) nonadiabatically localized in the up-field potential well is more favorable to be freed through the narrow and low inner-barrier when the molecule stretches to the critical internuclear distance. **(c)** Sketch of the angular streaking concept and our coordinate system. Due to the streaking of the elliptically polarized laser pulse (red helix), an electron will receive a final momentum (green arrows) oriented perpendicular to the laser field (violet arrows) at its tunneling. The instantaneous laser field vector at the moment of ionization hence can be retrieved from the streaked electron momentum $p_e$ (green arrows) or its recoil on the ion sum-momentum $p_{sum}$ (orange arrows). The olive helices in the dashed insets show sample momentum trajectories of the driven electron during the pulse. The ejection direction of the ionizing center or molecular orientation (orange blobs) is derived from the relative-momentum between the fragment ions $p_{rel}$ (blue arrows).

**Figure 2**

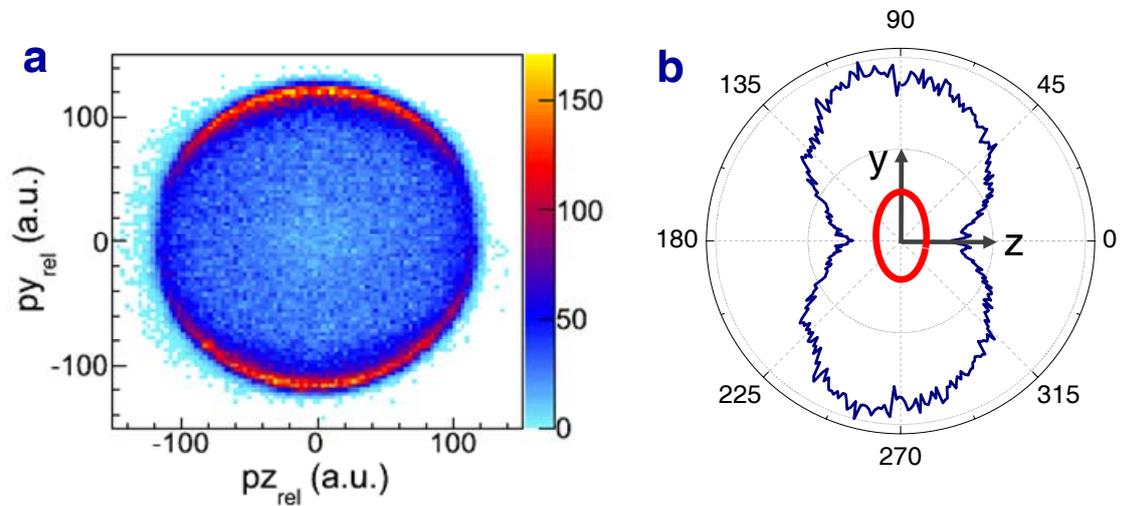

**Figure 2: Relative momentum and angular distributions of Ar$^+$+Xe$^+$ breakup**. **(a)** Measured relative momentum distribution between Ar$^+$ and Xe$^+$ from Coulomb exploding double ionization of ArXe using elliptically polarized femtosecond laser pulses whose major and minor axes orientated along *y*- and *z*-axis, respectively, as sketched by the red ellipse in **(b)**. **(b)** The corresponding angular distribution of the Ar$^+$+Xe$^+$ breakup in the polarization plane of the laser pulse. The double ionization rate maximizes for molecules orientated along the polarization major axis of our elliptically polarized laser field.

**Figure 3**

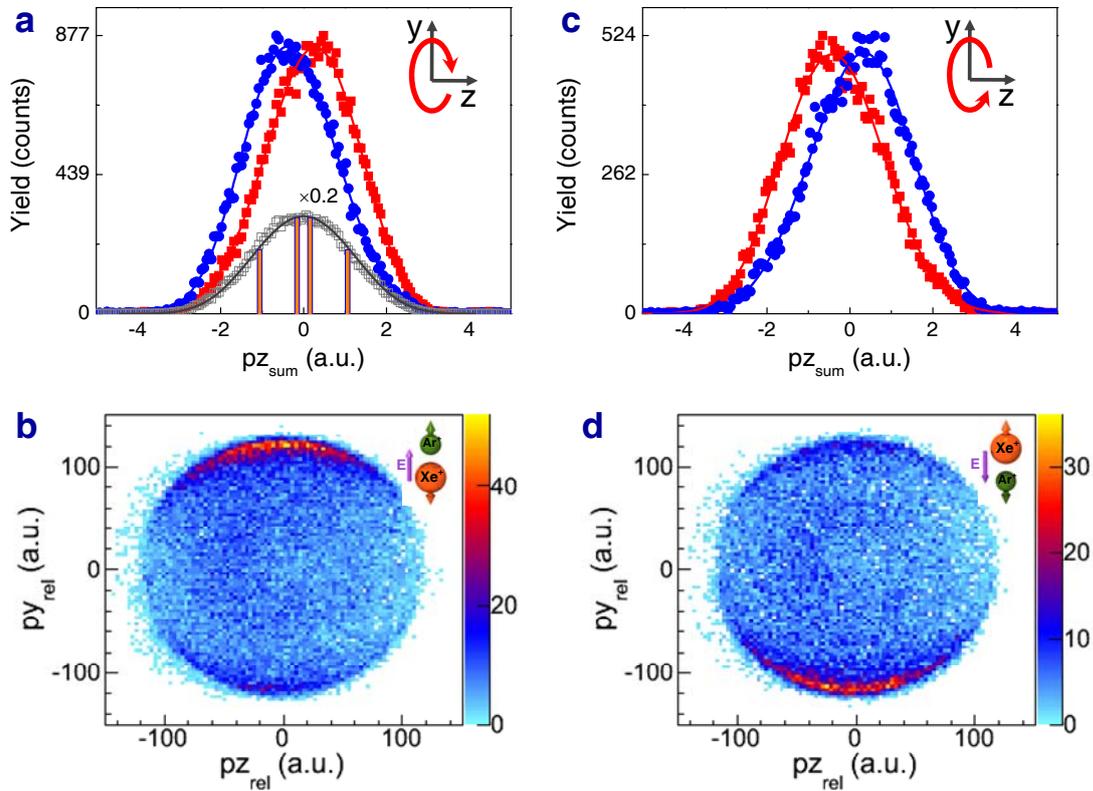

**Figure 3: Ion sum- and relative-momentum distributions of $Ar^+ + Xe^+$ breakup. (a,c)** Ion sum-momentum distributions. These reflect the recoil of two sequentially emitted electrons. The red squares and blue circles are events where $Ar^+$ fly in the $+y$ and $-y$ directions, and the gray squares are the sum of them. To guide the eye, four possible sum-momenta of the sequentially ionized electrons are indicated by the pink bars in **(a)** under the gray curve. **(b,d)** Ion relative-momentum distributions. These reflect the ArXe orientation. Only the events when the instantaneous field vector (violet arrow) at the release of the second electron pointing along **(b)** $+y$, or **(d)** $-y$ are selected by choosing $pz_{sum} > 1.0$ a.u. (see text for details). In agreement with the intuitive picture of the enhanced ionization, the second electron from ArXe is favored to be freed when the instantaneous laser field is pointing from $Xe^+$ towards the Ar atom. As sketched by the inset red ellipses at the top-right corners of **(a)** and **(c)**, the sense of rotation of the electric field vector of our laser pulse is adjusted to be **(a,b)** clockwise, and **(c,d)** anticlockwise, respectively.

**Figure 4**

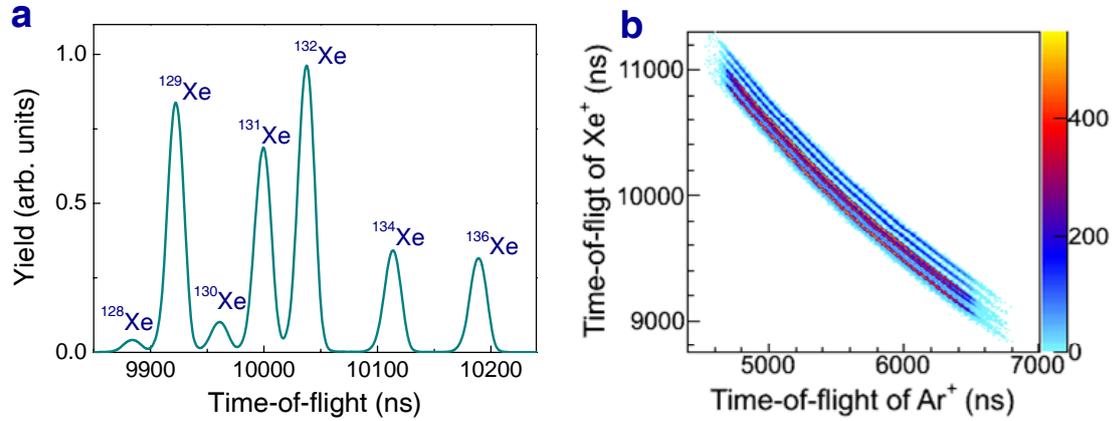

**Figure 4: Two-particle correlation identifies ArXe isotopes. (a)** Measured times-of-flight mass spectrum of the $Xe^+$ isotopes. **(b)** Correlations between the times-of-flight of the Coulomb exploding $Ar^+$ and $Xe^+$ following two-side double ionization of ArXe. Each sharp stripe stands for one ArXe isotope.

**Figure 5**

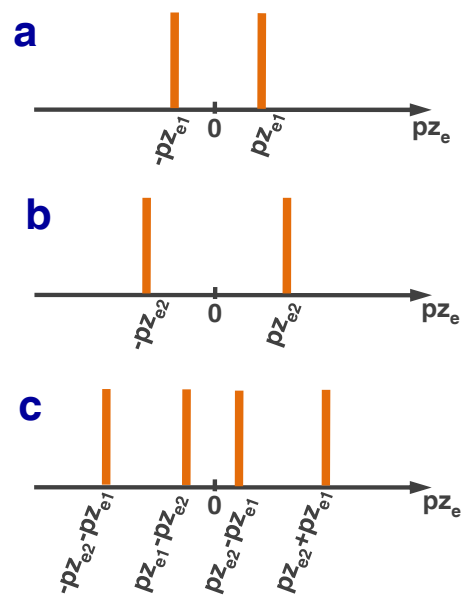

**Figure 5: Schematic illustration of the momentum convolution of two sequentially released electrons.** The orange bars indicate the momentum peaks of the **(a)** first and **(b)** second electrons, and **(c)** the eventually measured distribution as a result of their convolution.